\documentstyle[12pt,epsfig]{article}

\begin{document}

\begin{center}

{\baselineskip=24pt
{\Large {\bf Central Dominance and the Confinement Mechanism in Gluodynamics} }

{\baselineskip=16pt
\vspace{1cm}

{\large B.L.G.~Bakker$^a$, A.I.~Veselov$^b$, and M.A.~Zubkov$^b$ }\\

\vspace{.5cm}
{ \it
$^a$ Department of Physics and Astronomy, Vrije Universiteit,\\
De Boelelaan 1081, NL-1081 HV Amsterdam, The Netherlands

\vspace{1ex}

$^b$ ITEP, B.Cheremushkinskaya 25, Moscow, 117259, Russia }
 }
}
\end{center}

\vspace{3cm}

\begin{abstract}
\noindent
New topological objects, which we call center monopoles, naturally arise in the 
Maximal Center Projection of SU(3) gluodynamics. The condensate of the center
monopoles is the order parameter of the theory.
\end{abstract}

\vspace{1cm}

\newpage

\section{Introduction}
  The nonlinear structure of strong interaction physics explains
  why analytical approaches to the full investigation of the
  QCD internal structure are absent. Therefore QCD must
  be investigated by a combination of computer simulations
  and semi-analytical approaches.  Today our knowledge of gluodynamics
  includes several beautiful structures arising in the QCD vacuum.
  Among these structures we mention instantons,
  Maximal Abelian monopoles, center vortices, and gluonic strings.
   These objects influence the main physical effects such as
  asymptotic freedom, the creation of glueballs and confinement.

  In this work we investigate one of the structures mentioned above,
  corresponding to the center vortices in $SU(3)$
  gluodynamics (the $SU(2)$ case was investigated in \cite{su2}).
  This phenomenon was
  discovered by Greensite et al. (see, for example, \cite{Greensite}) and
  can be seen through the so-called Maximal Center Projection.  Within this
    picture it was  shown that string-like center vortices interact with
  quarks via topological Aharonov - Bohm forces. Moreover, these forces
  are responsible for confinement \cite{Tomboulis}.

It is already a tradition to describe confinement by the dual
  superconductor mechanism, which was proposed by Mandelstam.
    It seems that this mechanism is clearly
  realized in the $SU(2)$ simplification of QCD using the so-called
  Maximal Abelian projection \cite{Wiese}.  After this projection 
  monopoles appear, which are condensed in the confinement phase. Due
  to this condensation the color force lines are constricted into a
  string, which connects a quark and an antiquark. This string has nonzero
  tension, which leads to the confinement of quarks. A great number
  of physicists hope that the Maximal Abelian projection works in the
  same way in the real $SU(3)$ theory. We do not know if they are
  right or not, but in $SU(3)$ gluodynamics after the Maximal
  Abelian projection not one, but two monopoles appear, and even if the dual
  superconductor mechanism works here, it must be complicated
  \cite{Suzuki}.

  According to the Center Dominance hypothesis center vortices are
  responsible for confinement. 
Thus the following question arises in a natural way during the
investigation of the Maximal Center projection: What is the connection 
between center vortices and the dual superconductor picture.  
To answer this question we construct monopole-like objects from center 
vortices.  We call them
  center monopoles. The condensate of center monopoles is the order
  parameter. It is different from zero in the confinement phase of the
  $SU(3)$ theory. Thus one may expect that the center monopole
  is the monopole which plays a role in the dual Meissner mechanism.

\section{The Maximal Center Projection.}

We consider $SU(3)$ gluodynamics with  the Wilson action
$S(U) = \beta \sum_{\mathrm{plaq}} (1-1/3 \mathrm{Re} \, \mathrm{Tr} U_{\mathrm{plaq}})$. Here the sum is
over the plaquettes of the lattice. If the given plaquette consists
of the links $[xy]$,$[yz]$,$[zw]$,$[wx]$ then $U_{\mathrm{plaq}} = U_{[xy]}
 U_{[yz]}  U_{[zw]}  U_{[wx]} $.

The Maximal Center Projection makes the link
 matrix $U$ as close as possible to the elements of the center $Z_3$ 
of $SU(3)$: 
$ Z_3 = \{{\rm diag}(\mathrm{e}^{(2\pi i /3) N}, \mathrm{e}^{(2\pi i /3) N}, \mathrm{e}^{(2\pi i /3) N}\}$,
where $N \in \{1, 0, -1\}$.  In this work we use the
  so-called indirect version of the Maximal Center Projection. This
  procedure works as follows.

First, make the functional 
\begin{equation} 
Q_1 = \sum_{\mathrm{links}} (|U_{11}| + |U_{22}| + |U_{33}|)
\end{equation}
maximal with respect to the gauge transformations $U_{xy} \rightarrow g^{-1}_x
U_{xy} g_y$, thus fixing the Maximal Abelian gauge. As a consequence every link
matrix becomes almost diagonal.

Secondly, to make this matrix as close as possible to the center of $SU(3)$,
make the phases of the diagonal elements of this matrix maximally close to 
each other. This is done by minimizing the functional
\begin{eqnarray}
 Q_2 & = & \sum_{\mathrm{links}}[(1-\cos({\rm Arg}(U_{11})-{\rm Arg}(U_{22})))
                    +(1-\cos({\rm Arg}(U_{11})-{\rm Arg}(U_{33}))) \nonumber \\
 &&                    +(1-\cos({\rm Arg}(U_{22})-{\rm Arg}(U_{33})))].
\end{eqnarray}
with respect to the gauge transformations.
This gauge condition is invariant under the central subgroup $Z_3$ of $SU(3)$.

The center vortices are defined as follows.
After fixing the Maximal Center gauge we define the
integer-valued link variable $N$:

\begin{eqnarray}
N_{xy}=0 &{\rm if}& ({\rm Arg}(U_{11})+{\rm Arg}(U_{22})+{\rm Arg}(U_{33}))/3
 \in \; ]-\pi/3, \pi/3], \nonumber \\
N_{xy}=1 &{\rm if}& ({\rm Arg}(U_{11})+{\rm Arg}(U_{22})+{\rm Arg}(U_{33}))/3
 \in \; ]\pi/3, \pi], \nonumber \\
N_{xy}=-1 &{\rm  if}& ({\rm Arg}(U_{11})+{\rm Arg}(U_{22})+{\rm Arg}(U_{33}))/3
 \in \; ]-\pi, -\pi/3].
\end{eqnarray}

In other words $N=0$ if $U$ is close to $1$,  $N=1$ if $U$ is close to 
\noindent
$\mathrm{e}^{2\pi i/3}$ and  $N=-1$ if $U$ is close to $\mathrm{e}^{-2\pi i/3}$.

Next we define the plaquette variable:

\begin{equation}
 \sigma_{xywz} =  N_{xy}+N_{yw}-N_{zw}-N_{xz}
\end{equation}

We introduce the dual lattice and define the variable $\sigma^*$ dual to
 $ \sigma$: if plaquette $^*\Omega$ is dual to plaquette $\Omega$, then 
$\sigma^*_{^*\Omega} = \sigma_{ \Omega}$.  One can
easily check that the variable $\sigma$ represents a closed
surface. This surface  is known as the worldsheet of the center
vortex.

We express the $SU(3)$
gauge field $U$  as the product of $\exp((2\pi i / 3) N)$  and $V$,
where  $V$ is the $SU(3)/Z_3$ variable 
$({\rm Arg}(V_{11})+{\rm Arg}(V_{22})+{\rm Arg}(V_{33}))/3 \in \; ]-\pi/3,
\pi/3]$. Then
$U=\exp((2\pi i/ 3) N) V$

After that we represent the action of the Wilson loop $C$ as follows:
\begin{equation}
 W_C = \Pi_C U =\exp((2\pi i / 3) L(C,\sigma)) \Pi_C V
\end{equation}
The term $(2\pi i/3)
L(C,\sigma)$ is known as the Aharonov - Bohm interaction term.
The quantity $L(C,\sigma)$ is the linking number of the loop $C$ and the
closed surface $\sigma^*$.

The content of the center dominance hypothesis is that after the 
Maximal Center projection the Aharonov - Bohm interaction term by itself
causes confinement and produces the full string tension.

The center monopole is the $Z_3$ analogue of the monopole in $U(1)$
theory. Let us recall that monopoles in $U(1)$ theory
are constructed as loops on which the force lines of the
gauge field end. It is well known that in electrodynamics the Maxwell
equations $dF=0$ restrict the existence of magnetic charges. But
in the compact theory values of F which differ from each other
by $2\pi$, are equivalent. Thus the correct field strength is
$F \, {\rm mod} \, 2\pi$ and $^*d (F \, {\rm mod} \, 2\pi) = 2\pi j_m$, 
where $j_m$ is the monopole current.

The Aharonov - Bohm interaction between the center vortex and the quark depends
only on $[\sigma] \, {\rm mod} \, 3$.  Here $\sigma$ is the $Z_3$ analogue of 
the $U(1)$ field strength.
The variable $[\sigma] \, {\rm mod} \, 3$ represents the surface with boundary.
This boundary is a closed line. We assume that this line represents
the world trajectory of the particle, which we call a center monopole:
\begin{equation}
 3j_m = \; ^*d ([\sigma] \, {\rm mod} \, 3) =
 \delta  ([\sigma^*] \, {\rm mod} \, 3).  
\end{equation}
(Here we use the notations of differential forms on the lattice.
For the definition of these notations see, for example, \cite{forms}.)

We suggest the reader to consider the center monopole as the monopole
which condensation leads to formation of the quark-antiquark string
according to the dual superconductor mechanism. The results of the
next section partially justify this hypothesis. In particular, we find that
in the finite temperature theory the condensate of the center
monopoles
is the order parameter. The center monopoles are condensed indeed in
the confinement phase.

\section{Numerical results}

We used a lattice of size $16^3 \times 4$. The
confinement - deconfinement phase transition for this lattice is at
$\beta = 5.69$ approximately \cite{engels}. Our results are as follows

\begin{figure}[!htb]
\begin{center}
\begin{tabular}{cc}
\epsfig{file=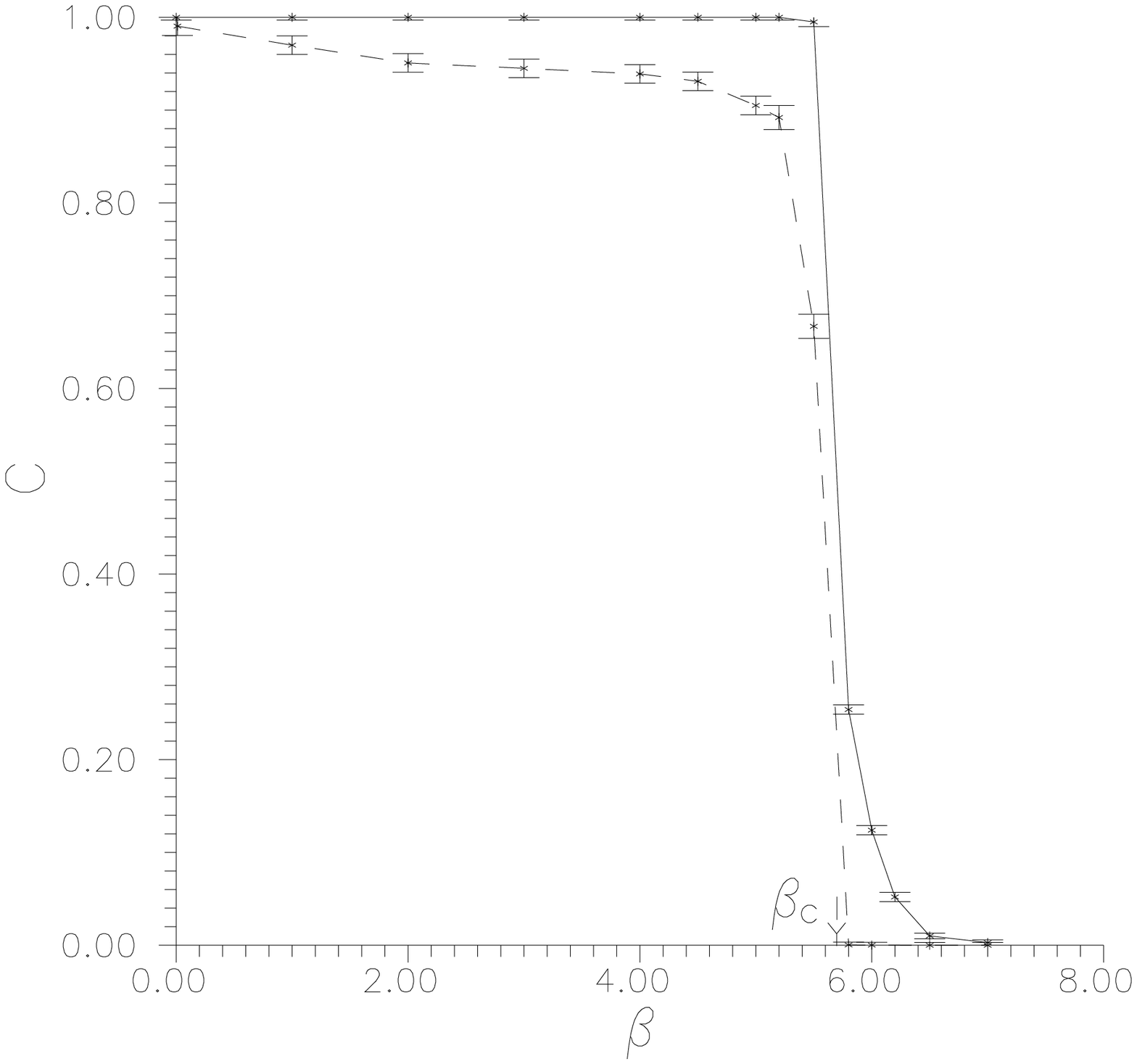,width=15.0cm,height=15.0cm,angle=0}\\
\end{tabular}
\end{center}
\caption[]{
Percolation properties of center vortices (solid curve) and center monopoles
 (dashed curve). \label{fig1}}
\end{figure}

\begin{figure}[!htb]
\begin{center}
\begin{tabular}{cc}
\epsfig{file=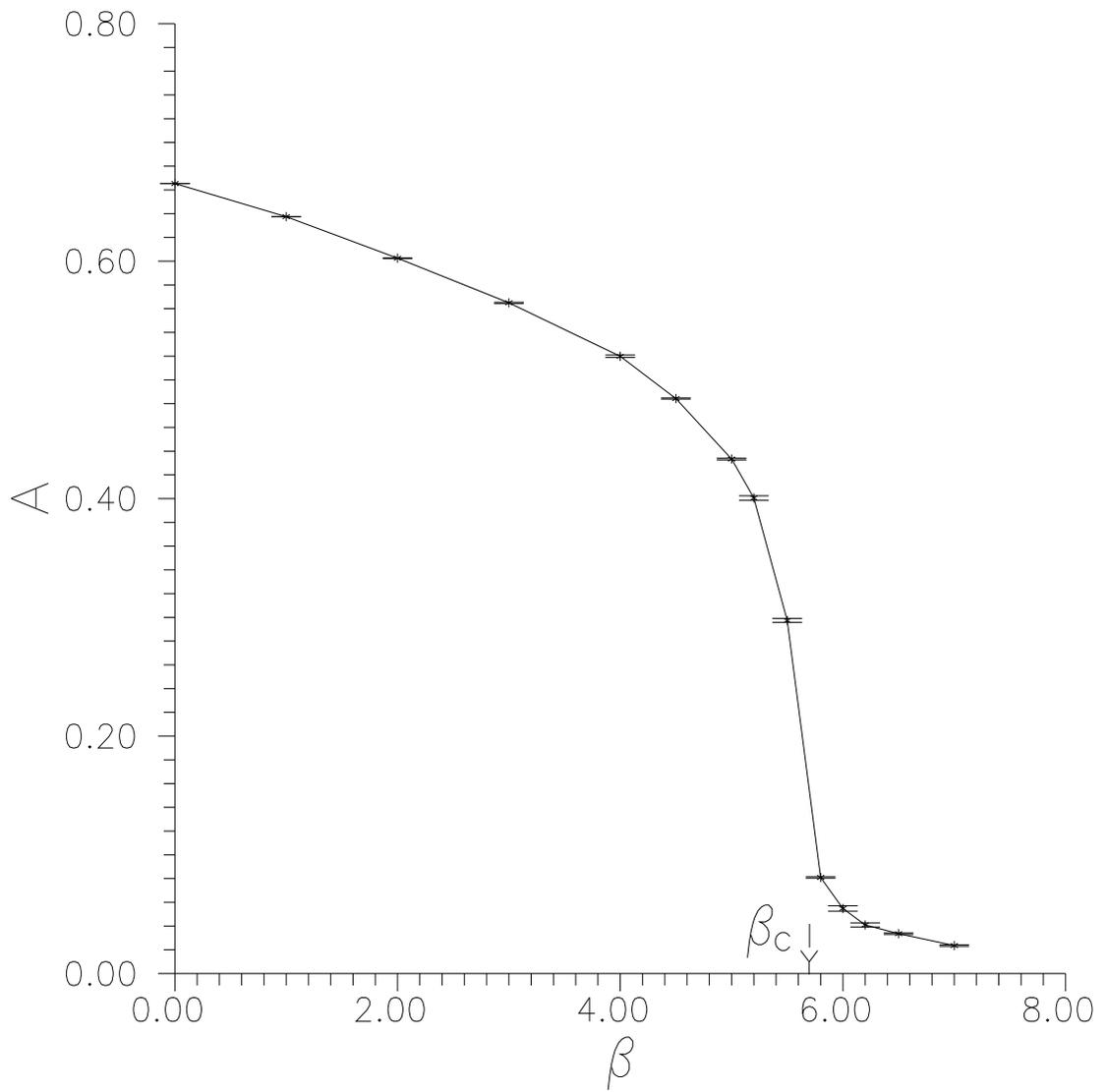,width=15.0cm,height=15.0cm,angle=0}\\
\end{tabular}
\end{center}
\caption[]{The density of the center vortices.\label{fig2}}
\end{figure}

\begin{figure}[!htb]
\begin{center}
\begin{tabular}{cc}
\epsfig{file=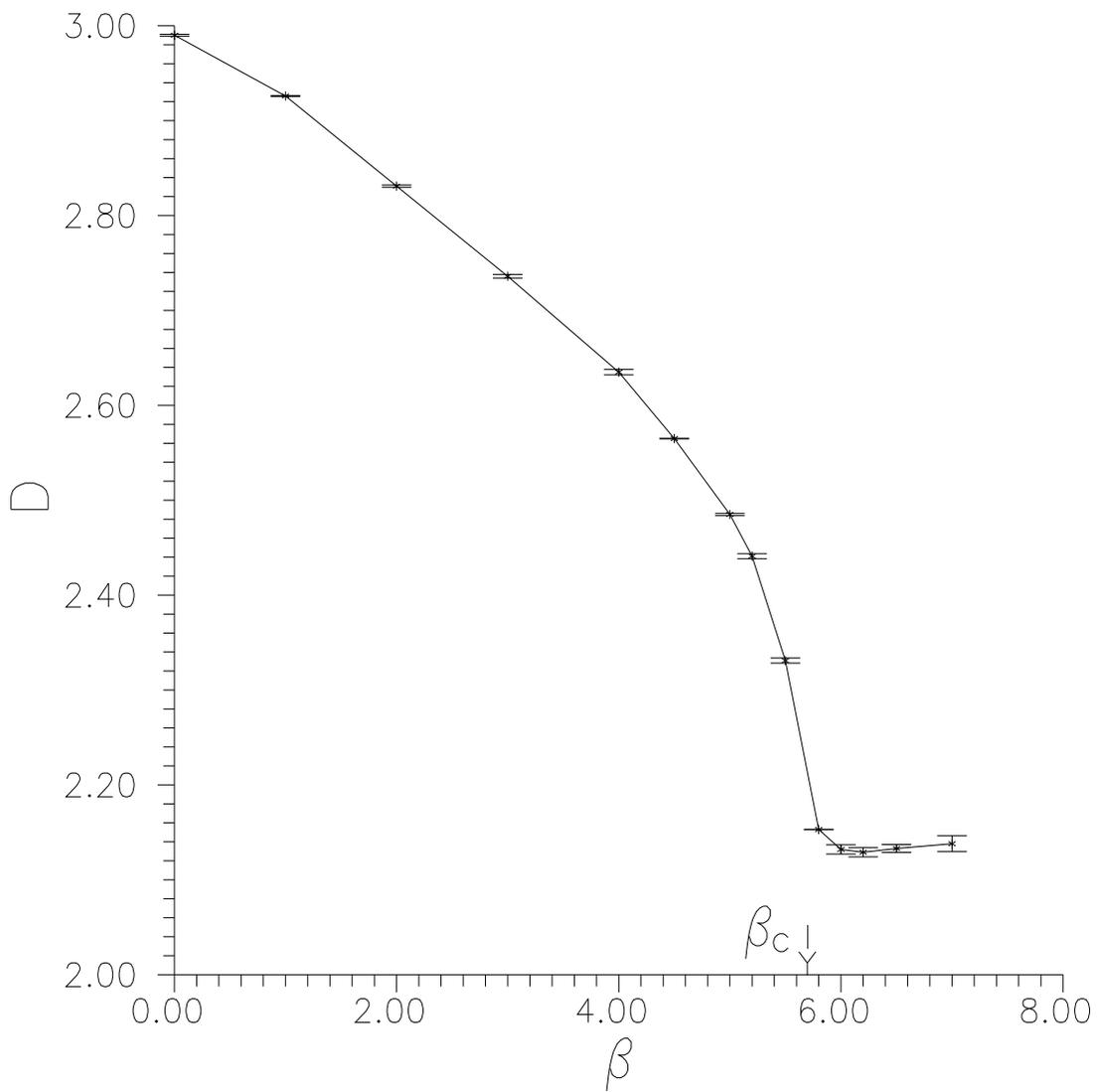,width=15.0cm,height=15.0cm,angle=0}\\
\end{tabular}
\end{center}
\caption[]{
The fractal dimension of the center vortices.\label{fig3}}
\end{figure}

\begin{enumerate}
\item  

The center vortices are condensed in
the confinement phase, and not condensed at high temperature.  This
  follows from the consideration of the probability that two points are
connected by the string  worldsheet. We find for this probability:  
$ \rho(x,y)_{\mathrm{vort}} \rightarrow C_{\mathrm{vort}}(\beta)$ at 
$|x-y|\rightarrow \infty$.  We observe, that
$C_{\mathrm{vort}}$ is equal to $1$ in the confinement phase, and falls to zero
in the deconfinement phase.  (The solid line in Fig.~\ref{fig1}.)

\item 
The density of the center vortices is represented in Fig.~\ref{fig2}.
\item 
The fractal dimension of center vortices  which is 
given by $D = 1+2A/L$, where $A$ is the number of plaquettes and $L$ is the
number of links on the string, is represented in Fig.~\ref{fig3}.
\item 
The center monopoles are
condensed in the confinement phase, but not condensed in the
deconfinement phase.  This follows from a consideration of the
probability that two points are connected by a monopole worldline. We
find for this quantity:  
$ \rho(x,y)_{mon} \rightarrow C_{mon}(\beta)$ at $|x-y|\rightarrow \infty$.  
We observe, that $C_{mon}$ is equal to $0$ in the deconfinement
phase, and different from $0$ in the confinement phase.
The condensate as a function of $\beta$ is represented in Fig.~\ref{fig1}
by the dashed line.
\item 
In addition we notice here, that according to the percolation
properties,   the center vortices and center monopoles are
distributed homogeneously. This follows from the fact that the
probability of two points to be connected by the worldsheet or
the worldline of these objects
 is constant for all distances for center monopoles and
center vortices.
\end{enumerate}

\section{Conclusions}

The Maximal Abelian Projection which has been used actively lately in the
investigation of strong interaction physics, leads to the existence
of two monopoles within SU(3) gauge theory. Thus the dual superconductor 
mechanism of confinement becomes complex and unnatural.  
In our work we use the Maximal Center
Projection and find that after applying this procedure a new
interesting object arises.  We call it the Center monopole. It turns out
that this monopole is condensed in the confinement phase and is not
condensed in the deconfinement phase.  Thus we believe this object
to be a good candidate for the monopole that works in the dual
superconductor mechanism.

\section*{Acknowledgments}

The authors are grateful to M. Chernodub, J. Greensite,
M. Polykarpov, T. Suzuki and and E. Tomboulis for useful discussions.
We are very grateful for SU(3) computer code given to us by S. Kitahara.
This work was supported by the JSPS Program on Japan--FSU scientists
collaboration, by the grants INTAS 96-370,
INTAS-RFBR-95-0681 and RFBR-97-02-17491.

\newpage

\end{document}